\documentclass[prl,showpacs,floats,twocolumn,a4paper]{revtex4}%
\usepackage{epsf}
\usepackage{graphicx}
\usepackage{amsmath}
\usepackage{amsfonts}
\usepackage{amssymb}
\usepackage{dcolumn}%
\providecommand{\U}[1]{\protect\rule{.1in}{.1in}}

\begin{document}
\title{Thermal Spin-Transfer Torques in Magnetoelectronic Devices}
\author{Moosa Hatami and Gerrit E. W. Bauer}
\affiliation{Kavli Institute of NanoScience, Delft University of Technology, Lorentzweg 1,
2628 CJ Delft, The Netherlands}
\author{Qinfang Zhang and Paul J. Kelly}
\affiliation{Faculty of Science and Technology and MESA+ Institute for Nanotechnology,
University of Twente, P.O. Box 217, 7500 AE Enschede, The Netherlands}
\date{\today}

\begin{abstract}
We predict that the magnetization direction of a ferromagnet can be reversed
by the spin-transfer torque accompanying spin-polarized thermoelectric heat
currents. We illustrate the concept by applying a finite-element theory of
thermoelectric transport in disordered magnetoelectronic circuits and devices
to metallic spin valves. When thermalization is not complete, a spin heat
accumulation vector is found in the normal metal spacer, \textit{i.e}., a
directional imbalance in the temperature of majority and minority spins.

\end{abstract}

\maketitle

Spintronics seeks to exploit the interplay of conduction electron charge and
spin transport in nanostructures. The modulation of an electric current in a
spin valve, \textit{i.e}. a normal-metal spacer sandwiched between two
ferromagnets, by the relative magnetization directions is the essence of the
celebrated giant magnetoresistance (GMR) effect. Magnetization reversal by the
current-induced spin-transfer torque in spin valves or tunnel junction
\cite{Slonczewski-Berger96,Stiles06} has already been applied in memory
devices as a low-power alternative to Oersted-field magnetization switching
\cite{Sony}.

Increasing data storage density and access rate is a continuing challenge for
the magnetic recording industry. The relatively high current densities and
voltages that are required to operate magnetic random access memories give
rise to heating effects that will complicate modeling and deteriorate device
stability and lifetime, making it difficult to reduce device sizes. Controlled
heating can, however, also be beneficial: recording by thermally assisted
reversal of magnetization via short laser pulses \cite{McDaniel05} or by Joule
heating at highly resistive thin layers \cite{Prejbeanu04} is a possible
solution for the next generation of high-density non-volatile data storage.

Johnson and Silsbee \cite{Johnson87} and Wegrowe \cite{Wegrowe00} analyzed the
thermodynamics of transport in collinear ferromagnetic-normal metal
heterostructures in the diffuse regime. The measured magnetothermoelectric
power and Peltier effect of multilayered nanowires has been described in terms
of spin-dependent and spin-flip processes in the bulk layers by Gravier
\textit{et al.} \cite{Gravier06}. A large Peltier cooling effect in transition
metal nanopillars has been measured by Fukushima \textit{et al}.
\cite{Fukushima05}. Nonlinear thermoelectric transport in non-collinear
magnetic tunnel junctions has been studied numerically in a tight-binding
approximation \cite{Wang01}. Tsyplyatyev \textit{et al}. invoked thermally
excited spin currents \cite{Tsyplyatyev06} to explain thermomagnetic effects
in metals with embedded ferromagnetic clusters \cite{Clusters}\textbf{.} In
this Letter we report another example of \textquotedblleft
spin-caloritronics\textquotedblright, \textit{viz}. a strong coupling of
thermoelectric spin and charge transport with the magnetization dynamics in
nanoscale magnetic structures.We establish the existence of a thermally
induced torque on the magnetization at an interface between a normal metal and
a ferromagnet. The conditions that should be met in order to observe this
phenomenon experimentally are discussed in detail for disordered transition
metal-based ferromagnet%
$\vert$%
normal metal%
$\vert$%
ferromagnet spin valves which opens the possibility to switch magnetizations
by \textit{e.g}. pulsed laser heating.

In bulk metallic systems, electron transport is well described by
semiclassical diffusion theory \cite{Jedema01-02}. However, atomically sharp
interfaces should be treated using quantum mechanical scattering matrices
\cite{Schep97}. In mesoscopic systems such as quantum point contacts
\cite{Molenkamp92} scattering theory is a well established framework for
understanding thermoelectric transport \cite{Butcher90}. Here we treat
magnetic nanostructures in terms of electronic distribution functions in
\textquotedblleft bulk\textquotedblright\ layers that are connected with
boundary conditions at interfaces in terms of microscopic scattering matrices
using an extension of a finite-element (or circuit) theory
\cite{Nazarov94,Brataas00,Brataas06} to treat charge, spin and energy currents
on an equal footing. Interface scattering is parameterized by a few
material-specific conductances that are accessible to first-principles calculations.

We start by partitioning a conducting structure into discrete low-resistance
nodes connected by resistive elements. Ferromagnetic (F) or normal metal (N)
nodes are characterized by $2\times2$ distribution matrices in spin space that
can be expanded into a scalar and a vector component $\hat{f}^{\mathrm{F}%
(\mathrm{N})}=f_{c}^{\mathrm{F}(\mathrm{N})}\hat{1}+\hat
{\mbox{\boldmath$\sigma$}}\cdot\mathbf{s}^{\mathrm{F}(\mathrm{N})}%
f_{s}^{\mathrm{F}(\mathrm{N})}.$ The unit vector of the spin quantization axis
$\mathbf{s}^{\mathrm{F}}$ is parallel to the magnetization of the ferromagnet,
whereas $\mathbf{s}^{\mathrm{N}}$ can point in any direction. An imbalance
between the distribution functions at two neighboring nodes induces a
non-equilibrium current. In linear response, the $2\times2$ spectral current
in spin space across a ferromagnet-normal metal junction at energy $\epsilon$
in the absence of spin-flip and inelastic scattering is given by Ohm's law
\cite{Brataas00}
\begin{equation}
\hat{\imath}_{\mathrm{N|F}}\left(  \epsilon\right)  =\sum_{\alpha\beta
}G^{\alpha\beta}\left(  \epsilon\right)  \hat{u}^{\alpha}\left[  \hat
{f}^{\mathrm{F}}\left(  \epsilon\right)  -\hat{f}^{\mathrm{N}}\left(
\epsilon\right)  \right]  \hat{u}^{\beta},
\end{equation}
where $\hat{u}^{\uparrow(\downarrow)}=(\hat{1}\pm\hat
{\mbox{\boldmath$\sigma$}}\mathbf{\cdot m})/2$ are projection matrices in
which $\hat{1}$ is the $2\times2$ unit matrix, the unit vector $\mathbf{m}$
denotes magnetization direction in the ferromagnet, and $\hat
{\mbox{\boldmath$\sigma$}}$ is the vector of Pauli matrices. The conductance
tensor elements read $G^{\alpha\beta}=\left(  e^{2}/h\right)  \sum_{nm}%
[\delta_{mn}-r_{nm}^{\alpha}(r_{nm}^{\beta})^{\ast}]$ in terms of the
energy-dependent reflection coefficients $r_{nm}^{\alpha}(\epsilon)$ for
majority and minority spins at the $\mathrm{N|F}$ interface. The total
charge-spin and heat matrix currents are defined as $\hat{I}=\int
d\epsilon\;\hat{\imath}(\epsilon)$ and $e\hat{\dot{Q}}=\int d\epsilon
(\epsilon-\mu)\hat{\imath}(\epsilon)=e\hat{I}^{\varepsilon}-\mu\hat{I},$
respectively, where $\mu$ is the equilibrium chemical potential and $\hat
{I}^{\varepsilon}$ the energy current. The charge and spin electric currents
$I_{c}$ and $\mathbf{I}_{s}$ are the scalar and vector components of the
matrix current $\hat{I}=(I_{c}\hat{1}+\hat{\mbox{\boldmath$\sigma$}}%
\cdot\mathbf{I}_{s})/2$. Analogously, $\hat{\dot{Q}}=(\dot{Q}_{c}\hat{1}%
+\hat{\mbox{\boldmath$\sigma$}}\cdot\mathbf{\dot{Q}}_{s})/2$.

When inelastic scattering in a given\ node is weak, the concept of a local
temperature is not applicable and the distribution function has to be
determined as a function of energy \cite{Pierre01}, as will be discussed in a
future publication. Here we assume either that the applied voltage is much
smaller than the temperature or that there is sufficient inelastic scattering
so that $f_{\uparrow(\downarrow)}=f_{c}%
\genfrac{}{}{0pt}{}{+}{\left(  -\right)  }%
f_{s}$ may be parameterized by Fermi-Dirac distribution functions with
spin-dependent chemical potentials $\mu_{\uparrow(\downarrow)}=\mu
-eV_{\uparrow(\downarrow)}$ and temperatures $T_{\uparrow(\downarrow)}$ that
are weakly perturbed from their values at equilibrium $\left(  \mu,T\right)
$\textbf{. }When conductances\ do not vary too rapidly in an energy interval
$k_{B}T$ around the Fermi level, Sommerfeld's expansion of the distribution
functions up to order $(k_{B}T/\mu)^{2}$ may be invoked, where $k_{B}$ is
Boltzmann's constant \cite{Ashcroft}. Defining charge and spin temperatures
$T_{c}=(T^{\uparrow}+T^{\downarrow})/2$ and $T_{s}=T^{\uparrow}-T^{\downarrow
}$ (similarly for $\mu_{c}$ and $\mu_{s}$, $V_{c}$ and $V_{s}$ ), we also
require $T_{s}\ll2T_{c}$ in the following. The Sommerfeld expansion leads to
integrals of the form $\int d\epsilon(\epsilon-\mu)^{d}f_{s}(\epsilon)$ that
for $d=0,1,2$ read $\mu_{s}$, $(\pi^{2}k_{B}^{2}/3)TT_{s}$ and $(\pi^{2}%
k_{B}^{2}/3)T^{2}\mu_{s}$, respectively. The same integrals over the function
$f_{c}^{\mathrm{F}}-f_{c}^{\mathrm{N}}$ result in similar expressions by
$\mu_{s}\rightarrow\mu_{c}^{\mathrm{F}}-\mu_{c}^{\mathrm{N}}$ and
$T_{s}\rightarrow T_{c}^{\mathrm{F}}-T_{c}^{\mathrm{N}}$. The spin and heat
currents through an $\mathrm{N|F}$ interface are spanned by longitudinal
components polarized parallel to $\mathbf{m}$\textbf{\ (}$I_{s}^{\Vert
}=\mathbf{m\cdot I}_{s}$ and $\dot{Q}_{s}^{\Vert}=\mathbf{m\cdot\dot{Q}}_{s}%
$\textbf{\ )} and transverse contributions $\mathbf{I}_{s}^{\bot}%
=\mathbf{I}_{s}-I_{s}^{\Vert}\mathbf{m}$ and $\mathbf{\dot{Q}}_{s}^{\bot
}=\mathbf{\dot{Q}}_{s}-\dot{Q}_{s}^{\Vert}\mathbf{m.}$ \begin{widetext}
The matrix that relates
the particle, heat, and spin currents is equivalent to those found in the
literature \cite{Johnson87,Gravier06} when $\mathbf{m}\parallel\mathbf{s}$:
\begin{equation}
\left(
\begin{array}
[c]{c}%
I_{c}\\
\dot{Q}_{c}\\
I_{s}^{\Vert}\\
\dot{Q}_{s}^{\Vert}%
\end{array}
\right)  =G\left(
\begin{array}
[c]{cccc}%
1 & -S & P & -P^{\prime}S\\
-ST & L_{0}T & -P^{\prime}ST & PL_{0}T\\
P & -P^{\prime}S & 1 & -S\\
-P^{\prime}ST & PL_{0}T & -ST & L_{0}T
\end{array}
\right)  \left(
\begin{array}
[c]{c}%
-(V_{c}^{\mathrm{F}}-V_{c}^{\mathrm{N}})\\
T_{c}^{\mathrm{F}}-T_{c}^{\mathrm{N}}\\
-(V_{s}^{\mathrm{F}}-\mathbf{m}\cdot\mathbf{s}V_{s}^{\mathrm{N}})/2\\
(T_{s}^{\mathrm{F}}-\mathbf{m}\cdot\mathbf{s}T_{s}^{\mathrm{N}})/2
\end{array}
\right)  ,\label{par}%
\end{equation}
where $G=G^{\uparrow}+G^{\downarrow}$ is the total conductance, $S=-eL_{0}%
T\partial_{\epsilon}\ln G|_{\epsilon_{F}}$ is the thermopower (Mott's law),
both at the Fermi energy ($\epsilon_{F}$) and $L_{0}=(\pi k_{B}/e)^{2}%
/3=2.45\times10^{-8}\mathrm{W}\Omega\mathrm{K}^{-2}$ is the Lorenz number.
$P=\left(  G^{\uparrow}-G^{\downarrow}\right)  /G$ is the polarization of the
conductance with $\left\vert P\right\vert \leq1$ and $P^{\prime}%
=\partial_{\epsilon}\left(  G^{\uparrow}-G^{\downarrow}\right)  |_{\epsilon
_{F}}/\partial_{\epsilon}G|_{\epsilon_{F}}$ is the polarization of its energy
derivative at the Fermi energy. In contrast to $P$, $|P^{\prime}|$ is not
bounded and $P^{\prime}S$ can be very large when a van Hove singularity is
close to the Fermi energy for one spin direction.
$P_{S}=\left(  S_{\uparrow}-S_{\downarrow}\right)  /\left(  S_{\uparrow
}+S_{\downarrow}\right)  =\left(  P^{\prime}-P\right)  /\left(  1-P^{\prime
}P\right)  $ is the spin polarization of the thermopower. We focus here on the
transverse spin currents:
\begin{equation}
\left(
\begin{array}
[c]{c}%
\mathbf{I}_{s}^{\bot}\\
\mathbf{\dot{Q}}_{s}^{\bot}%
\end{array}
\right)  =\left(
\begin{array}
[c]{cc}%
\operatorname{Re}G^{\uparrow\downarrow}\mathbf{m\times}+\operatorname{Im}%
G^{\uparrow\downarrow}\; & eL_{0}T\left(  \operatorname{Re}G_{\epsilon
}^{\uparrow\downarrow}\mathbf{m\times}+\operatorname{Im}G_{\epsilon}%
^{\uparrow\downarrow}\right)  \\
eL_{0}T^{2}\left(  \operatorname{Re}G_{\epsilon}^{\uparrow\downarrow
}\mathbf{m\times+}\operatorname{Im}G_{\epsilon}^{\uparrow\downarrow}\right)
& L_{0}T\left(  \operatorname{Re}G^{\uparrow\downarrow}\mathbf{m\times
+}\operatorname{Im}G^{\uparrow\downarrow}\right)
\end{array}
\right)  \left(
\begin{array}
[c]{c}%
V_{s}^{\mathrm{N}}\mathbf{s\times m}\\
-T_{s}^{\mathrm{N}}\mathbf{s\times m}%
\end{array}
\right)  ,\label{perp}%
\end{equation}
\end{widetext}which are parameterized by the (spin-)mixing conductance
$G^{\uparrow\downarrow}$ and its energy derivative $G_{\epsilon}%
^{\uparrow\downarrow}=\partial_{\epsilon}G^{\uparrow\downarrow}|_{\epsilon
_{F}}$. We disregard the imaginary part of the mixing conductance
\cite{Brataas06} and its energy derivative. In analogy with the dimensionless
mixing conductance $\eta=2\mathrm{\operatorname{Re}}G^{\uparrow\downarrow}/G$
we also introduce a dimensionless \textquotedblleft mixing
thermopower\textquotedblright\ as $\eta^{\prime}=2\mathrm{\operatorname{Re}%
}G_{\epsilon}^{\uparrow\downarrow}/G_{\epsilon}$. Both transverse spin
currents $\mathbf{I}_{s}^{\bot}$ and $\mathbf{\dot{Q}}_{s}^{\bot}$ are
absorbed by the ferromagnet and transferred as a torque on the magnetization
order parameter.

We extend the methodology used to calculate bare interface conductances at the
Fermi energy \cite{Xia16} to obtain its energy dependence $G^{\alpha\beta
}(\epsilon)$. A finite drift is taken into account by replacing $G^{\alpha
\beta}(\epsilon)^{-1}$ with $G^{\alpha\beta}(\epsilon)^{-1}-\left(
h/2e^{2}\right)  \left(  N^{\mathrm{N}}(\epsilon)^{-1}+\delta_{\alpha\beta
}N^{\alpha\mathrm{F}}(\epsilon)^{-1}\right)  $ \cite{Schep97,Brataas06}, where
$N^{\alpha F}(\epsilon)$ is the number of propagating modes of spin $\alpha
$\ at energy $\epsilon\ $in \textrm{F}. To determine the thermopower $\ln
G^{\alpha\beta}(\epsilon)$ is differentiated numerically. The results for $S$,
$P^{\prime},$ $P_{S}$, $\eta$ and $\eta^{\prime}$ are listed for a number of
important interfaces in Table I. Note that the spin polarization of the
thermopower in bulk magnets, believed to be dominated by electron-magnon
spin-flip scattering \cite{Fert92,Gravier06}, has a different origin.
\begin{table}[ptb]
\begin{ruledtabular}
\begin{tabular}
{l rD{.}{.}{-0} D{.}{.}{-0} D{.}{.}{-0} D{.}{.}{-2}  D{.}{.}{-2}}
& \multicolumn{1}{c}{ $ \frac{S} {T}  ({\rm nV/K^2}) $}
& \multicolumn{1}{c}{ $P^{\prime}(\%)$}
& \multicolumn{1}{c}{ $P_S$(\%)}
& \multicolumn{1}{c}{ $\eta$}
& \multicolumn{1}{c}{ $\eta^{\prime}$ }   \\
\hline
Cu/Co(001)  & -13    &   72 &    -8 & 0.50   &  0.03  \\
Cu/Co(001)* & -34    &   89 &    43 & 0.49   &  0.06  \\
\hline
Cu/Co(110)  & -10    &    6 &   -66 & 0.67   & -0.32  \\
Cu/Co(110)* & -13    &   85 &    45 & 0.63   &  0.07  \\
\hline
Cu/Co(111)  & -15    &   56 &    -6 & 0.53   &  0.13  \\
Cu/Co(111)* & -15    &   77 &    17 & 0.64   &  0.13  \\
\hline
Cr/Au(001)  &   7    &    0 &     0 &$-$     & $-$    \\
Cr/Au(001)* &   0    &    0 &     0 &$-$     & $-$    \\
\hline
Cr/Fe(001)  &  22    &  -40 &    48 & 4.23   & -4.27  \\
Cr/Fe(001)* &   7    & -190 & -9500 & 3.25   & -0.48  \\
\hline
Cr/Co(001)  &  62    & -111 &  -160 & 3.03   & -2.86  \\
Cr/Co(001)* &  23    &  -95 &   -92 & 2.92   & -0.86  \\
\end{tabular}
\end{ruledtabular}
\caption{Thermoelectric interface parameters calculated at the Fermi energy
for a number of almost lattice-matched interfaces including a drift correction
\cite{Schep97,Brataas06}. The star * indicates a dirty interface modeled in a
10x10 lateral supercell with two layers of 50\%-50\% alloy. }%
\end{table}

The temperature $T_{c}$, voltage $V_{c}$, particle spin-accumulation
$\mathbf{s}V_{s}$ and temperature spin-accumulation $\mathbf{s}T_{s}$ of a
given node are governed by Kirchhoff rules. Charge and angular momentum
conservation implies that the sum of all charge and spin currents flowing into
a given node vanishes. Since thermal transport in metals is dominated by the
conduction electrons \cite{Gundrum05} we disregard the phonon contribution to
the energy currents. Electrons experience inelastic electron-electron and
electron-phonon collisions. $T_{s}$ decays by the energy exchange between the
electrons. The spin accumulation $V_{s}$\ is dissipated to the lattice by
spin-flip scattering which can be very weak in selected metals and is
disregarded here for simplicity. We distinguish two different regimes by
comparing the dwell time $\tau_{d}=e^{2}\mathcal{D}/\left(  4\pi G\right)  ,$
where $\mathcal{D}$ is the density of states, with $\tau_{E},$ the energy
relaxation time: electrons are completely thermalized when $\tau_{E}\ll
\tau_{d},$ but effectively non-interacting in the opposite regime $\tau_{E}%
\gg\tau_{d}$. The electron dwell time in metallic nanopillars with a spacer
thickness of 10 nm can be estimated to be $\sim100$ fs. At low temperatures
this can be much shorter than either electron-electron or electron-phonon
scattering times \cite{Pothier97} and the spin temperature difference or spin
heat accumulation becomes an important parameter. For elevated temperatures
inelastic scattering is more effective and we adopt a complete thermalization
model\textbf{.}

\begin{figure}[t]
\resizebox{\columnwidth}{!} {\includegraphics{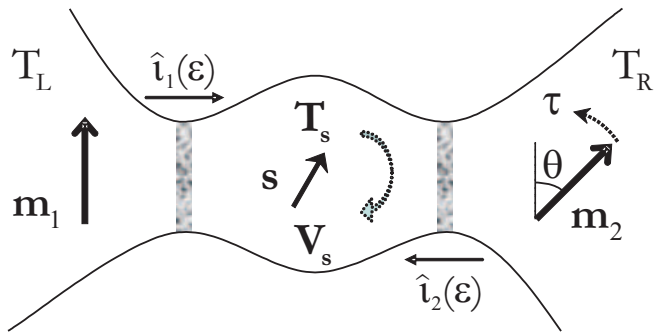}}
\caption{Schematic of a noncollinear $F(\mathbf{m}_{1})|N|F(\mathbf{m}_{2})$
spin-valve. Spin-dependent resistive elements separate the ferromagnetic
reservoirs and the normal metal node. A temperature bias induces a spin
accumulation in the form of heat and angular momentum imbalance, the interplay
of which is governed by inelastic scattering.}%
\label{fig1}%
\end{figure}

We illustrate the theory for symmetric $\mathrm{F}_{L}(\mathbf{m}%
_{1})|\mathrm{N}|\mathrm{F}_{R}(\mathbf{m}_{2})$ spin valves (see
Fig.~\ref{fig1}) consisting of two ferromagnetic reservoirs separated by a
normal metal node via two resistive contacts with variable magnetization
directions. We calculate the electric particle and heat currents and the
spin-transfer torques for a voltage bias $\Delta V=V_{R}-V_{L}$ and
temperature bias $\Delta T=T_{R}-T_{L}$ in the thermalized as well as
non-interacting regimes.  

In the \textit{thermalized }regime the spin heat accumulation $T_{s}%
^{\mathrm{N(F)}}=0$. In the steady state, lattice and electron temperature are
the same and energy is conserved in the node's electronic system. We find for
the total electric current as a function of the angle $\theta$ between the two
magnetizations
\begin{equation}
I_{c}=\frac{G}{2}\left(  \Delta V+S\Delta T\right)  -\frac{PG}{2}\frac
{\tan^{2}\theta/2}{\eta+\tan^{2}\theta/2}\left(  P\Delta V+P^{\prime}S\Delta
T\right)  . \label{Ic}%
\end{equation}
The angular magnetoresistance for $\Delta T=0,$ measured by Urazhdin
\textit{et al}. \cite{Urazhdin05}, is well described by circuit theory
\cite{Kovalev06}. When the current bias vanishes, a temperature bias $\Delta
T$ induces an angular magnetothermopower $\Delta V$ that depends on both $P$
and $P^{\prime}$
\begin{equation}
\left(  \frac{-\Delta V}{S\Delta T}\right)  _{I=0}=\frac{\eta+(1-PP^{\prime
})\tan^{2}\theta/2}{\eta+(1-P^{2})\tan^{2}\theta/2}%
\end{equation}
The angular dependence of the heat current
\begin{align}
&  \dot{Q_{c}}=-\Pi\frac{G}{2}(\Delta V+S\Delta T)-\frac{\kappa}{2}\Delta
T\nonumber\\
&  +\Pi\frac{P^{\prime}G}{2}\frac{\tan^{2}\theta/2}{\eta+\tan^{2}\theta
/2}(P\Delta V+P^{\prime}S\Delta T)
\end{align}
where $\Pi=ST$ is the interface Peltier coefficient, strongly violates the
Wiedemann-Franz law ($\kappa\approx L_{0}TG$). \textbf{ }A non-negative
entropy production rate in the Sommerfeld approximation requires
$|S_{max}|=\sqrt{L_{0}}\simeq157%
\operatorname{\mu V}%
/%
\operatorname{K}%
$ \cite{Guttman95}\textbf{\textbf{.}}

The spin-transfer torque exerted on the magnetizations by a temperature
difference over the spin valves in the thermalized electron regime reads
($\tau=\tau_{\Delta V}+\tau_{\Delta T}$)
\begin{equation}
\tau=\frac{G}{2}\frac{\eta\sin\theta}{\eta(1+\cos\theta)+(1-\cos\theta
)}\left(  P\Delta V+P^{\prime}S\Delta T\right)  . \label{tau}%
\end{equation}
We can understand the similarity of the torque induced by the voltage and
temperature bias as follows. A temperature difference over the spin valves
initially induces different temperatures for the spin species in the normal
metal node. Since we consider here the strongly interacting regime, such a
temperature difference relaxes quickly due to collisions that exchange energy
between spin systems but conserve the total energy. This is possible only by
generating a spin current and accumulation that subsequently induces a torque
just as the voltage does.

The dynamics of the magnetic layers is governed by a Landau-Lifshitz-Gilbert
equation augmented by the spin-transfer torque. We use Slonczewski's estimate
for the critical current \cite{Slonczewski-Berger96} that leads to
magnetization reversal in metallic ferromagnets, replacing $P\Delta V$ with
$P^{\prime}S\Delta T$.\textbf{ }A thermoelectric voltage $S\Delta
T\sim100\mathrm{%
\operatorname{\mu V}%
}$ corresponds to typical switching current densities of $10^{7}\mathrm{%
\operatorname{A}%
\operatorname{cm}%
}^{-2}$. Assuming that a laser pulse provides local heating corresponding to
$\Delta T\sim100\mathrm{K}$, we require $P^{\prime}S\sim1\mathrm{%
\operatorname{\mu V}%
/}\mathrm{%
\operatorname{K}%
}$, which is not an unrealistic value at room temperature (see Table I). When
the magnetic layers become thicker, the bulk resistance and thermopower of the
layers dominate. \textbf{ }The series resistor rule $S/G\simeq\sum_{i}%
S_{i}/G_{i},$ where $G_{i}$ and $S_{i}$ account for both bulk layers and
interfaces in a multilayer structure, holds for $P_{S_{i}}\ll1.$ Using the
bulk parameters by Gravier\textit{ et al}. \cite{GravierFukushima06} we
estimate that the effective thermopower can be much higher than a $\mathrm{%
\operatorname{\mu V}%
/}\mathrm{%
\operatorname{K}%
}$, implying a strongly increased relative efficiency of thermal magnetization
reversal $\tau_{\Delta T}/\tau_{\Delta V}$ for thicker magnetic layers. The
conditions for thermal spin-torque switching are presumably more easily met in
spin valves based on magnetic semiconductors \cite{Ohno04}.

In the absence of energy relaxation in the normal node , a vector spin heat
accumulation $\mathbf{s}T_{s}^{\mathrm{N}}$ develops. When $S^{2}\ll L_{0}$,
the \textquotedblleft non-interacting\textquotedblright\ thermal spin-transfer
torque $\tau_{\Delta T}^{\ast}$ reduces to the simple expression:
\begin{equation}
\left.  \frac{\tau_{\Delta T}^{\ast}-\tau_{\Delta T}}{\tau_{\Delta T}%
}\right\vert _{\Delta V=0}=\frac{(\eta^{\prime}-\eta)P}{\eta P^{\prime}}%
\frac{\tan^{2}\theta/2}{\eta+\tan^{2}\theta/2}, \label{taustar}%
\end{equation}%
\begin{equation}
\left.  \frac{T_{s}^{\mathrm{N}}}{\Delta T}\right\vert _{\Delta V=0}%
=\frac{\eta P^{\prime}}{\left(  \eta^{\prime}-\eta\right)  \sin\theta/2}%
\frac{\tau_{\Delta T}^{\ast}-\tau_{\Delta T}}{\tau_{\Delta T}}.
\end{equation}

In conclusion, we presented a circuit theory of thermoelectric transport in
non-collinear spin-valves. In thinly layered structures, transport properties
are governed by interface conductances and their energy derivatives that have
been computed from first principles. We predict a spin-transfer torque
associated with purely thermal currents that can be large enough to reverse
magnetizations. The concepts of spin heat accumulation and spin-mixing
thermopower have been introduced to describe the thermoelectric transport in
different energy relaxation regimes. We expect that a temperature gradient can
excite magnetization dynamics in magnetic tunnel junctions and domain walls in
ferromagnetic wires as well.

We acknowledge helpful discussions with A. Brataas, Y. Tserkovnyak, A.
Fukushima, Y. Suzuki, S. Yuasa, A. Deac, X. Waintal, and H. Pothier. This work
has been supported by NanoNed, EC Contracts IST-033749 \textquotedblleft
DynaMax\textquotedblright\ and NMP-505587-1{}\textquotedblleft
SFINX\textquotedblright, the \textquotedblleft Stichting voor Fundamenteel
Onderzoek der Materie\textquotedblright\ (FOM) and the \textquotedblleft
Stichting Nationale Computer Faciliteiten\textquotedblright\ (NCF).

\end{document}